# Bulk Superconductivity Induced by Se Substitution in $BiCh_2$-Based Layered Compounds $Eu_{0.5}Ce_{0.5}FBiS_{2-x}Se_x$

Yosuke Goto, Ryota Sogabe, and Yoshikazu Mizuguchi*

Department of Physics, Tokyo Metropolitan University, Hachioji 192-0397, Japan

*E-mail: mizugu@tmu.ac.jp

**Abstract**

We report the effect of Se substitution on the crystal structure and superconductivity of $BiCh_2$-based (Ch: S, Se) layered compounds $Eu_{0.5}Ce_{0.5}FBiS_{2-x}Se_x$ ($x$ = 0–1). Crystal structure analysis showed that both lattice constants, $a$ and $c$, increased with increasing $x$, which is different from the related La-doped system, such as $Eu_{0.5}La_{0.5}FBiS_{2-x}Se_x$. This is due to Se substitution for both in-plane and out-of-plane Ch sites in the present Ce-doped system. Zero resistivity was observed for $x$ = 0.2–1 above 2 K. Superconducting properties of $Eu_{0.5}Ce_{0.5}FBiS_{2-x}Se_x$ was investigated by magnetic susceptibility, and the highest superconducting transition temperature of 3.5 K was obtained for $x$ = 0.6 with a large shielding volume fraction. The emergence of bulk superconductivity and metallic conductivity can be qualitatively described in terms of increased in-plane chemical pressure effect. A magnetic anomaly below 8 K, probably because of ferromagnetic order of magnetic moment of $Ce^{3+}$ ions, coexists with bulk superconductivity in $BiCh_2$ layer. Since the effect of Se substitution on the magnetic transition temperature is ignorable, we suggest that the coupling between the magnetic order at the (Eu,Ce)F layer and superconductivity at the $Bi(S,Se)_2$ layer is weak.

## Introduction

There has been considerable interest in the study of $BiCh_2$-based (Ch: S, Se) layered compounds because of their unique functionalities including superconductivity, thermoelectricity, and so forth [1]. The layered crystal structure of the $BiCh_2$-based system consists of conducting $BiCh_2$ layers alternating with electrically insulating (blocking) layers, which resembles those of cuprate or Fe-based high-transition-temperature ($T_c$) superconductors [2,3]. Since the discovery of superconductivity in $Bi_4O_4S_3$ and $REO_{1-x}F_xBiS_2$ (RE: rare-earth element), several types of $BiCh_2$-based superconductors have been reported [1,4,5]. Although many experimental and theoretical studies have been devoted to clarify the superconductivity mechanism and to increase the $T_c$ of the $BiCh_2$-based superconductors, the consensus on the nature of superconductivity is yet to be achieved. Furthermore, recent measurements of laser-based angle-resolved photoemission spectroscopy [6] and first-principles calculations [7] have demonstrated that weak-coupling electron-phonon mechanism cannot describe the emerging $T_c$ of the $BiCh_2$-based superconductors. Therefore, systematic characterization is still crucial to clarify the superconductivity mechanism of these compounds.

As an example, $LaOBiS_2$ is a semiconductor with a band gap value of 1 eV, and superconductivity is induced by $F^-$-substitution for $O^{2-}$. This result suggests the conventional scenario that the emergence of superconductivity owing to electron carrier doping into the conduction band consisting of hybridized Bi 6p and Ch p orbitals [8,9]. Although electron carrier doping is believed to be essential to induce superconductivity in $BiCh_2$-based compounds, optimally doped $LaO_{0.5}F_{0.5}BiS_2$ only shows weak (filamentary) superconductivity rather than bulk superconductivity. To induce bulk superconductivity in this system, external pressure effects [10–17] and/or element substitution at the La or S sites [18–21], are available. These characteristics have been qualitatively described by sufficient orbital overlapping between in-plane Bi and Ch as a results of in-plane chemical pressure effect [22]. Indeed, it has been reported that such pressure effects successfully suppressed the in-plane disorder of $BiCh_2$ conducting layer, which was demonstrated by means of synchrotron X-ray diffraction, extended X-ray absorption fine structure, and neutron diffraction [22−26]. The typical example is the $Eu_{0.5}La_{0.5}FBiS_{2-x}Se_x$ system [27]. $Eu_{0.5}La_{0.5}FBiS_2$, whose carrier concentration corresponds to 0.5 electron doping per Bi atoms as in the case of $LaO_{0.5}F_{0.5}BiS_2$, does not show bulk superconductivity, while the superconductivity is induced by Se substitution owing to the increased in-plane chemical pressure effect.

In the present study, we report the synthesis, crystal structure analyses, and superconducting properties of $Eu_{0.5}Ce_{0.5}FBiS_{2-x}Se_x$. It has been reported that $Eu_{0.5}Ce_{0.5}FBiS_2$ showed the coexistence of superconductivity and magnetic order of RE moment [28]. However, bulk superconductivity has yet been achieved for the compound. We have demonstrated Se substitution successfully induced bulk superconductivity in $Eu_{0.5}Ce_{0.5}FBiS_{2-x}Se_x$, which was qualitatively described by an increase of in-plane chemical pressure, as the case of La-doped $Eu_{0.5}La_{0.5}FBiS_{2-x}Se_x$. Although magnetic order due to the $Ce^{3+}$ moment coexisted with superconductivity in $BiCh_2$ layer, it was suggested that coupling between them was ignorable.

## Experiments

Polycrystalline samples of $Eu_{0.5}Ce_{0.5}FBiS_{2-x}Se_x$ with $x$ = 0, 0.2, 0.4, 0.6, 0.8, and 1 were prepared by solid-state reaction of EuS (99.9%), $Ce_2S_3$ (99.9%), $BiF_3$ (99.9%), Bi (99.999%), S (99.99%), and Se (99.999%). Stoichiometric mixtures of these starting materials were pressed into pellets and sealed in a sealed quartz tube. The heating temperature was optimized for each composition: the samples were heated for 20 h at 780 °C for $x$ = 0, at 700 °C for $x$ = 0.2 and 0.4, at 650 °C for $x$ = 0.6 and 0.8, and at 600 °C for $x$ = 1. The obtained product was ground, mixed, pelletized, and heated again under the same heating condition.

The obtained samples were characterized using powder X-ray diffraction (XRD) using a CuK$\alpha$ radiation (Rigaku, Miniflex 600 equipped with a D/teX Ultra detector). The XRD patterns were analyzed using the Rietveld method using RIETAN-FP code [29]. Crystal structure was depicted using VESTA software [30]. Magnetic susceptibility ($\chi$) as a function of temperature ($T$) was measured using a superconducting quantum interference device (SQUID) magnetometer (Quantum Design, MPMS-3) with an applied field of 5 Oe after both zero-field cooling (ZFC) and field cooling (FC). The $T$ dependence of electrical resistivity ($\rho$) was measured using a four-terminal method.

## Results and Discussion

Figure 1(a) shows the XRD patterns of $Eu_{0.5}Ce_{0.5}FBiS_{2-x}Se_x$ ($x$ = 0–1). Almost all of diffraction peaks are assigned to those of $EuFBiS_2$-type tetragonal phase, indicating that the obtained samples are nearly single-phase. Several diffraction peaks due to impurity phase such as $Bi_2Se_3$ were also observed, as indicated by the asterisks in the Figure 1(a). However, both of lattice constants $a$ and $c$ increase with increasing $x$, as shown in Figure 1(b), indicating Se ions are successfully and systematically substituted for S ions in $Eu_{0.5}Ce_{0.5}FBiS_{2-x}Se_x$. Notably, linear relationships between lattice constants and Se content were observed, which were different from the related La-doped systems, such as $Eu_{0.5}La_{0.5}FBiS_{2-x}Se_x$, where $a$ increases with increasing $x$, but $c$ does not show apparent changes with increasing $x$ [22]. These characteristics were most likely due to site selectivity of Se substitution. In the case of $Eu_{0.5}La_{0.5}FBiS_{2-x}Se_x$, Se ions selectively occupy in-plane Ch1 site rather than out-of-plane Ch2 site, where schematic representation of crystal structure was shown in Figure 1(c). Such selective substitution for Ch site resulted in anisotropic elongation of lattice constants of $Eu_{0.5}La_{0.5}FBiS_{2-x}Se_x$. On the other hand, as shown in Figure 2, substituted Se ions occupy both of the Ch sites (Ch1 and Ch2) for $Eu_{0.5}Ce_{0.5}FBiS_{2-x}Se_x$, resulting in increases of both $a$ and $c$ with increasing $x$. Figures 3(a)–(d) show the in-plane Bi–Ch1 distance, inter-plane Bi–Ch1 bond distance, Bi–Ch2 bond distance, and RE–Ch2 distance. These bond distances seem to be correlating with the lattice constants $a$ and $c$. For example, Bi–Ch1 and RE–Ch2 distance tend to increase with increasing $x$. Besides, Bi–Ch2 distance and inter-plane Bi–Ch1 distance seem to be decreased as a result of Se-substitution. Such shrinkage is different from the crystal structure evolution of $Eu_{0.5}La_{0.5}FBiS_{2-x}Se_x$ by Se substitution [31].

As previously demonstrated, the in-plane chemical pressure (CP) of $BiCh_2$-based layered compounds can be qualitatively described with the following equation: CP = ($R_{Bi}$ + $R_{Ch1}$)/(Bi–Ch1 bond distance), where $R_{Bi}$ is obtained from the average Bi–S bond distance for $LaO_{0.54}F_{0.46}BiS_2$ [22]. The Bi–Ch1 bond distance was calculated using the results of the Rietveld refinement. Figure 4 shows the in-plane CP as a function of $x$ for $Eu_{0.5}Ce_{0.5}FBiS_{2-x}Se_x$. The CP monotonically increase with increasing $x$, indicating orbital overlap between Bi and Ch is enhanced by Se substitution. However, the value is lower than that of La-substituted systems $Eu_{0.5}La_{0.5}OBiS_{2-x}Se_x$ [27]. This is because of indistinct site selectivity of Se substitution in $Eu_{0.5}Ce_{0.5}FBiS_{2-x}Se_x$, as described above.

Figure 4(a) shows $\chi$–$T$ plots with an applied magnetic field of 5 Oe after ZFC for $Eu_{0.5}Ce_{0.5}FBiS_{2-x}Se_x$. Clear diamagnetic signals were not obtained for $x = 0$. Besides, diamagnetic signals corresponding to the emergence of superconductivity were observed for $x \geq 0.2$. The largest shielding volume fraction and $T_c$ of 3.5 K were obtained for $x = 0.6$, as shown in Figure 4(b), indicating that this is a bulk superconductor. The large shielding fraction exceeding $-1/4\pi$ should be due to demagnetization effect for the polycrystalline sample containing many plate crystals. For $x = 0$, a gradual increase of $\chi$ was observed below 8 K. It has been reported that magnetic moment of $Ce^{3+}$ ions of $CeO_{0.3}F_{0.7}BiS_2$ aligned ferromagnetically, which was demonstrated using neutron scattering measurements [32]. Therefore, the increase of $\chi$ below 8 K in the present sample would be attributable to the ferromagnetic ordering of $Ce^{3+}$ moment, as well. It is interesting to note that the amplitude of the magnetic transition was suppressed for $x = 0.2$ and 0.4, and was emerged again for $x \geq 0.6$, as shown in Figure S2. Although superconductivity was suppressed with increasing $x$ for $x > 0.6$, the magnetic transition was almost unchanged in this region. In addition, the magnetic transition temperature never changes with $x$, while superconducting $T_c$ clearly depends on $x$. These results suggest that the coupling between superconductivity at the $BiCh_2$ layer and the magnetic order at the $Eu_{0.5}Ce_{0.5}F$ layer is notably weak. At present, we have no idea to explain the suppression of the amplitude of the magnetic transition for $x = 0.2$ and 0.4. However, the valence states of Eu and Ce in $Eu_{0.5}Ce_{0.5}FBiS_{2-x}Se_x$ would be essentially sensitive to its chemical composition and/or crystal structure, as recently reported in $Eu_{0.5}La_{0.5}FBiS_{2-x}Se_x$ [33]. In addition, X-ray absorption spectroscopy revealed that mixed valence of Ce ions was suppressed by F doping for $CeO_{1-x}F_xBiS_2$ [34,35]. It was suggested that Ce-S-Bi coupling channel is essential to drive the system from the valence fluctuation regime to the Kondo-like regime. Further investigation of local crystal/electronic structure is required to clarify the detailed relationship between magnetism and superconductivity in $Eu_{0.5}Ce_{0.5}FBiS_{2-x}Se_x$.

Figure 5 shows $\rho$ as a function of $T$. For $x \leq 0.4$, $\rho$–$T$ plots shows semiconducting-like behavior, that is, $\rho$ increases with decreasing $T$. This semiconducting-like $T$ dependence is probably due to localization of doped electrons. Metallic-like $\rho$–$T$ characteristics were observed for $x \geq 0.6$. Note that carrier concentration is believed to be almost identical in $Eu_{0.5}Ce_{0.5}FBiS_{2-x}Se_x$ because of isovalent substitution of $S^{2-}$ with $Se^{2-}$ ions [36]. However, metallic-like $\rho$–$T$ plots were induced by Se substitution owing to an increase of in-plane CP, as in the case of $Eu_{0.5}La_{0.5}FBiS_{2-x}Se_x$ [27].

On the superconducting characteristics, the onset of the superconducting transition was evaluated to be $T_c^{onset} = 2.3$ K for $x = 0$, although zero resistivity was not obtained down to 1.9 K, as shown in Figure 5(b). Besides, zero resistivity was observed for $x \geq 0.2$, which is consistent with a diamagnetic signal of $\chi$–$T$ plots. The $T_c^{zero}$ of 3.5 K was obtained for $x = 0.6$–1, as shown in Figure 5(c), and it was also consistent with $T_c$ estimated from $\chi$–$T$ plots. Notably, $\rho$ gradually decreases well above $T_c^{zero}$ : $T_c^{onset}$ can be estimated to be 5.4 K for $x = 0.6$. Such a gradual decrease of $\rho$ is probably because of a local evolution of superconductivity with a higher $T_c$, as demonstrated in $NdO_{1-x}F_xBi_{1-y}S_2$ single crystals [37].

A superconductivity phase diagram established using $\chi$–$T$ and $\rho$–$T$ plots are depicted in Fig. 7. The semiconducting-like $\rho$–$T$ behavior and filamentary weak superconducting transition with $T_c^{onset}$ of 2.3 K was observed $x = 0$. The semiconducting-like behavior was suppressed by Se substitution, and bulk superconductivity with a large shielding volume fraction was observed. The highest $T_c$ was 3.5 K for $x = 0.6$. Evidently, the superconducting properties of $Eu_{0.5}La_{0.5}OBiS_{2-x}Se_x$ was optimized for $x = 0.6$, at the boundary of semiconducting-metallic phase transition. Such emergence of metallic-like $\rho$–$T$ behavior and bulk superconductivity was quite similar to the case of related compounds, such as $Eu_{0.5}La_{0.5}FBiS_{2-x}Se_x$ [27]. Thus, the importance of the in-plane CP for the emergence of metallic conductivity and bulk superconductivity in $BiCh_2$-based compounds are clearly shown with the new $BiCh_2$-based superconductor system.

## Conclusions

We have synthesized a new $BiCh_2$-based superconductor system, $Eu_{0.5}Ce_{0.5}FBiS_{2-x}Se_x$. Polycrystalline samples for $x$ = 0–1 were prepared by a solid-state reaction. Both lattice constants, $a$ and $c$, monotonically increases with increasing $x$. This linear relationship between lattice constants and Se content is different from that of related compounds, such as $Eu_{0.5}La_{0.5}FBiS_{2-x}Se_x$. Crystal structure analysis by the Rietveld refinement showed that this is due to the indistinct site selectivity of Se substitution; namely, Se occupied both of in-plane Ch1 and out-of-plane Ch2 sites. The sample with $x$ = 0 exhibited the semiconducting-like $\rho$–$T$ and weak superconducting signals. Metallic-like conductivity and bulk superconductivity were induced by Se substitution. Such characteristics were qualitatively described in terms of increased in-plane CP, as in the case of related compounds. The highest $T_c$ was 3.5 K for $x$ = 0.6, and superconductivity phase diagram was established. Magnetic transition, probably due to ferromagnetic order of magnetic moment of $Ce^{3+}$, coexists with bulk superconductivity in $BiCh_2$ layer. The coupling between superconductivity and magnetic order is ignorable in $Eu_{0.5}Ce_{0.5}FBiS_{2-x}Se_x$.

## Acknowledgement

We thank to O. Miura of Tokyo Metropolitan University for the experimental support. This work was partly supported by Grant-in-Aid for Scientific Research (Nos. 15H05886, 16H04493, 17K19058, and 16K17944) and JST-CREST (No. JPMJCR16Q6), Japan.

Lett. 106, 67002 (2014).

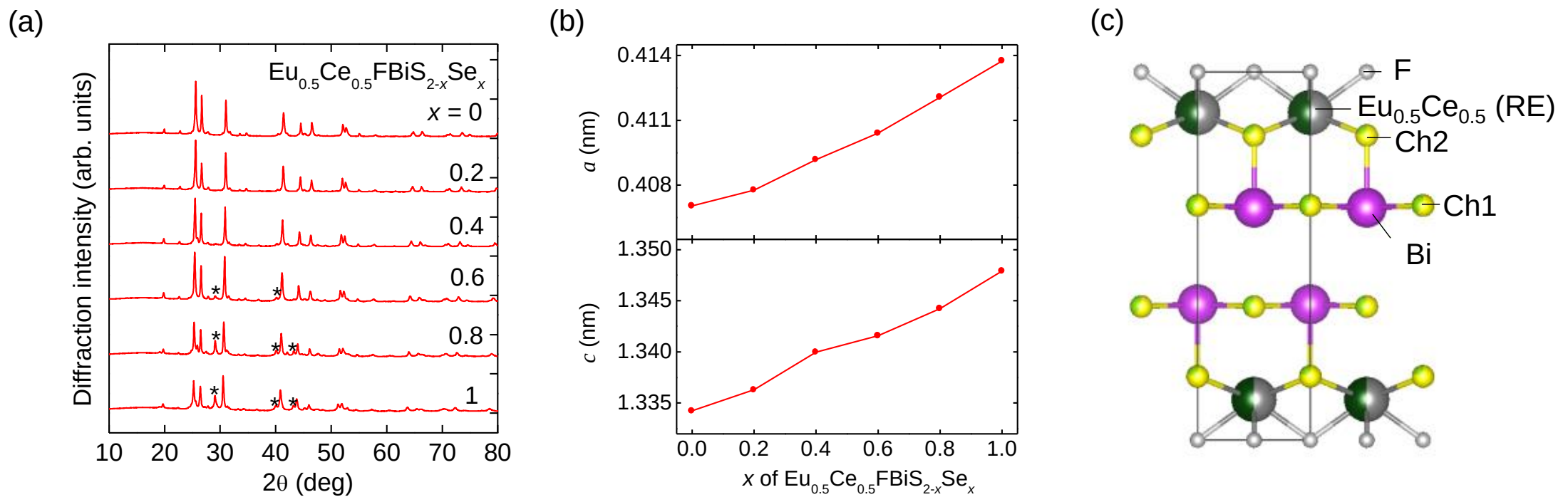

**Figure 1.** (a) XRD patterns of $Eu_{0.5}Ce_{0.5}FBiS_{2-x}Se_x$ for $x$ = 0–1. Asterisks denote the diffraction due to $Bi_2Se_3$ impurity phase. Rietveld fitting patterns were shown in Supplemental Materials. (b) Lattice constants versus $x$ for $Eu_{0.5}Ce_{0.5}FBiS_{2-x}Se_x$. (c) Schematic representation of crystal structure for $x$ = 0.6 as a representative sample.

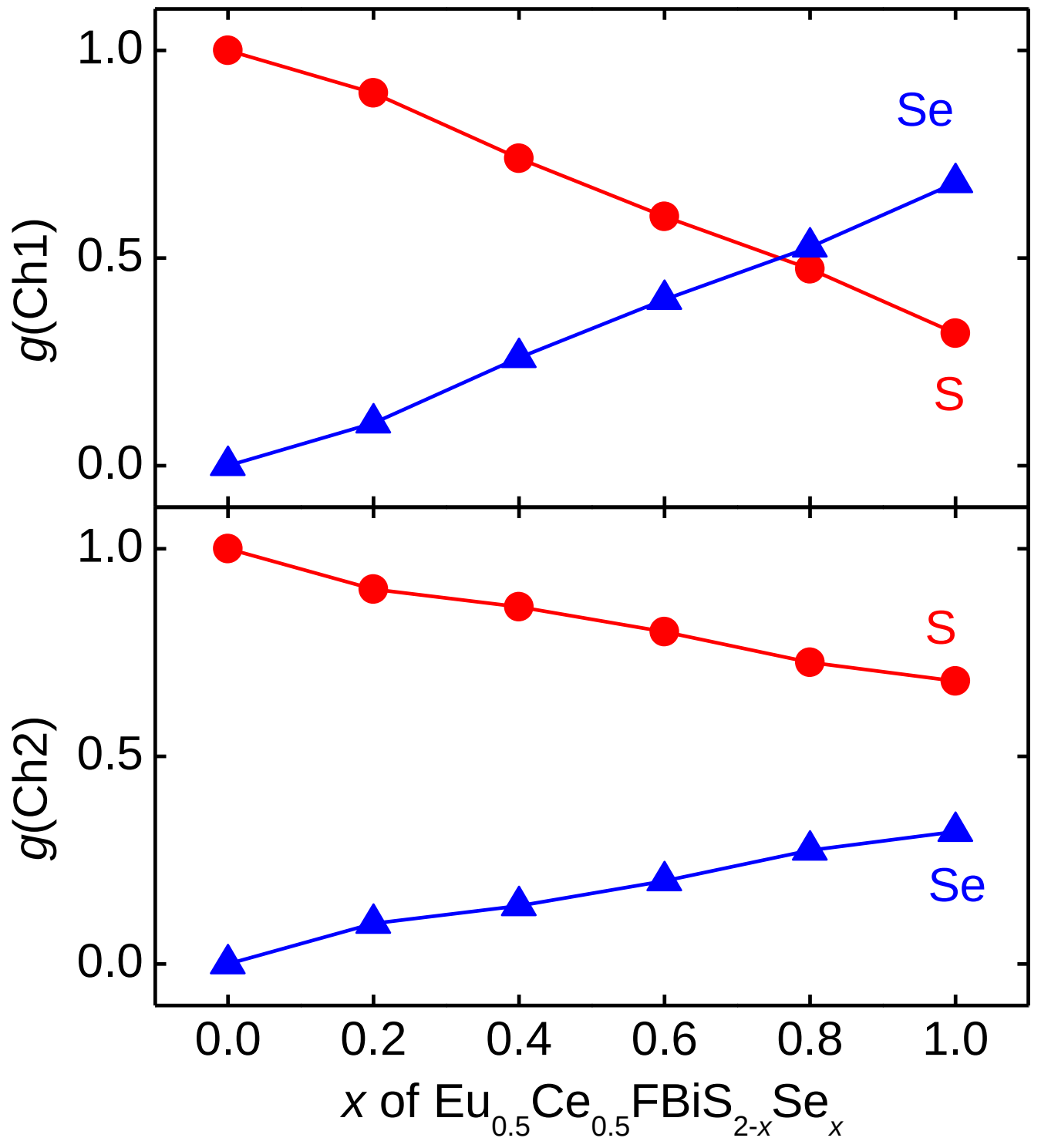

**Figure 2.** Site occupancies ($g$) of in-plane Ch1 and out-of-plane Ch2 sites for $Eu_{0.5}Ce_{0.5}FBiS_{2-x}Se_x$.

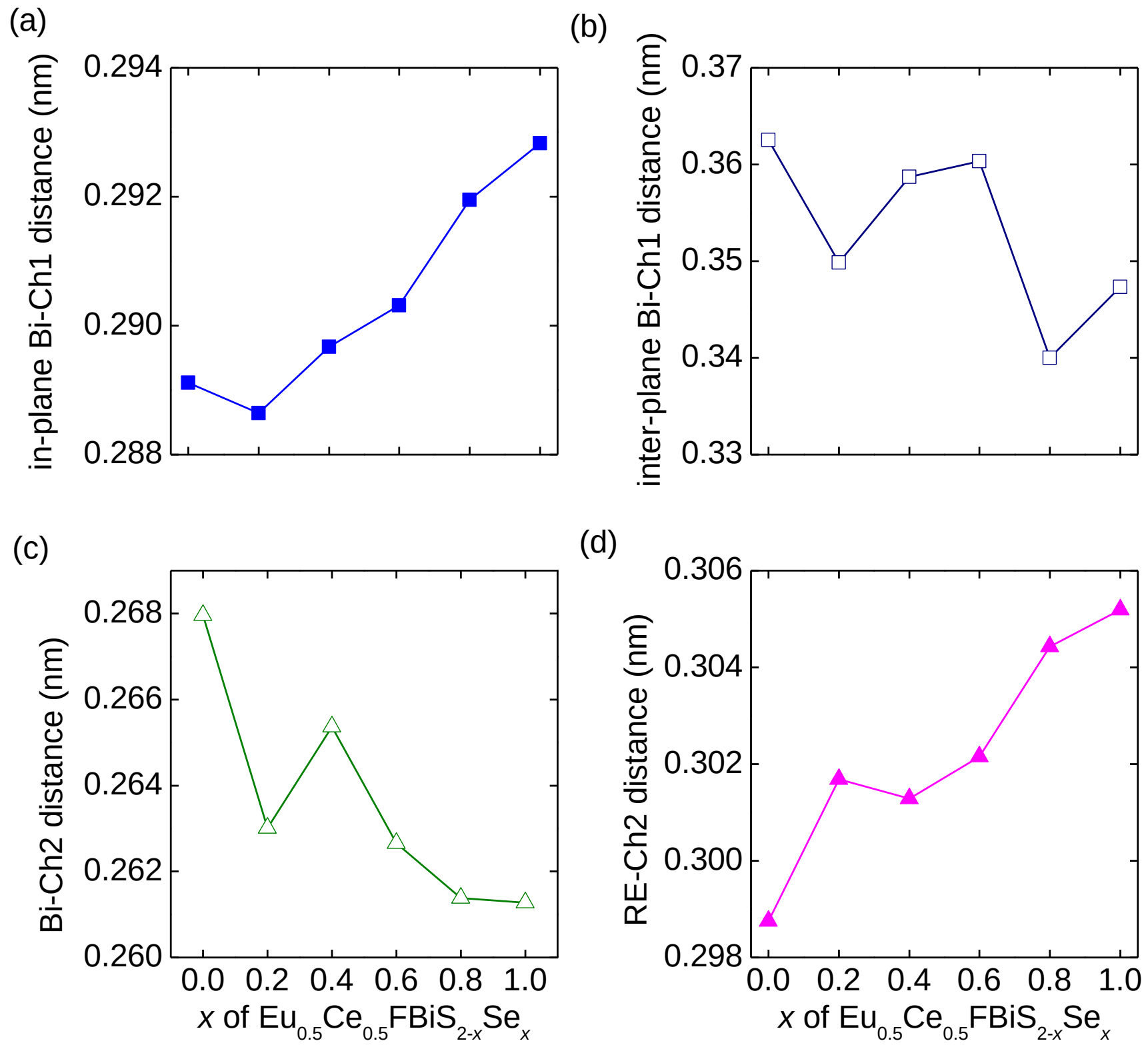

**Figure 3.** Se concentration dependences of selected bond distances for $Eu_{0.5}Ce_{0.5}FBiS_{2-x}Se_x$: (a) in-plane Bi-Ch1 distance, (b) inter-plane Bi-Ch1 distance, (c) Bi-Ch2 distance, and (d) RE-Ch2 distance.

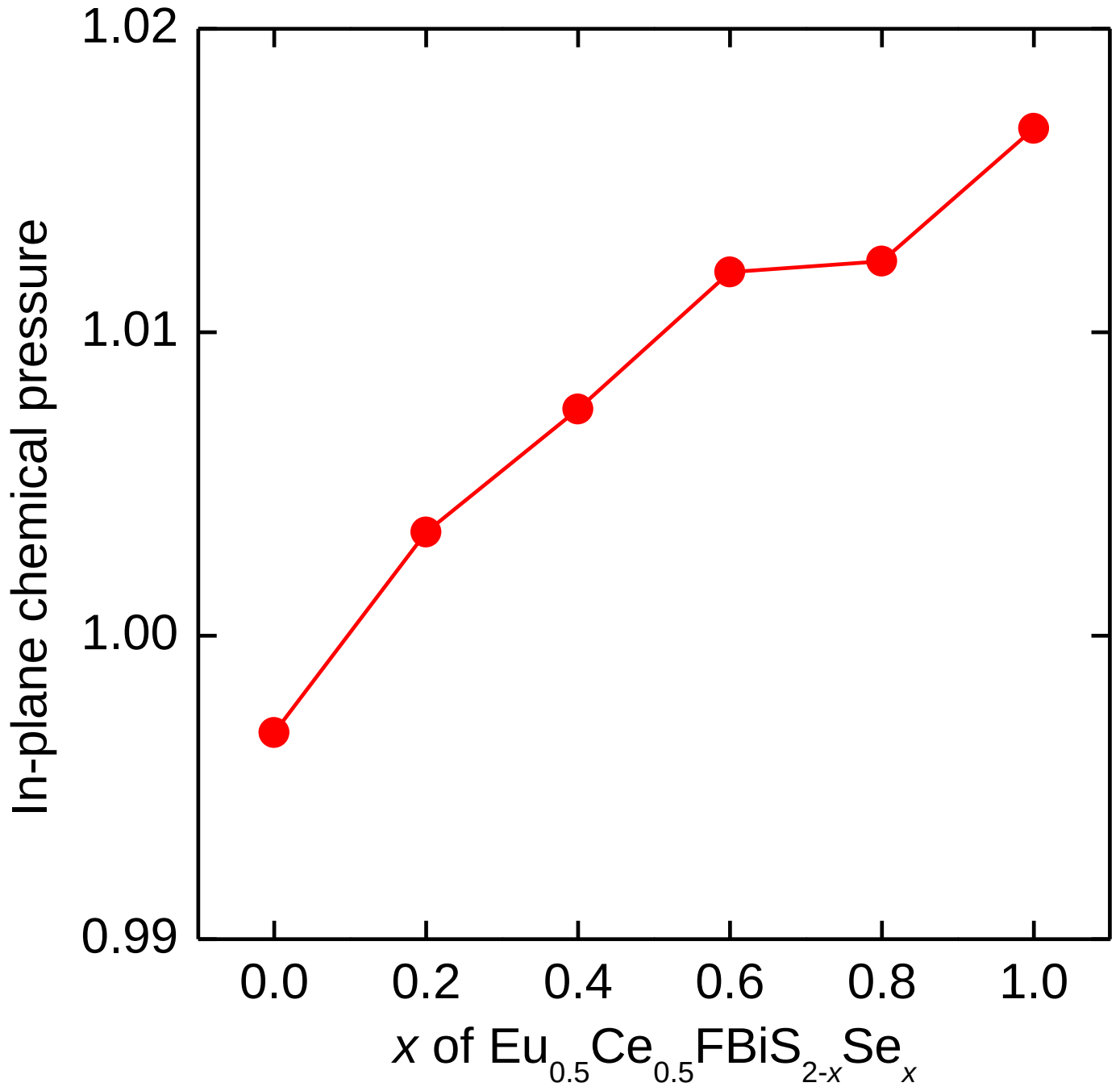

**Figure 4.** In-plane chemical pressure as a function of Se content for $Eu_{0.5}Ce_{0.5}FBiS_{2-x}Se_x$.

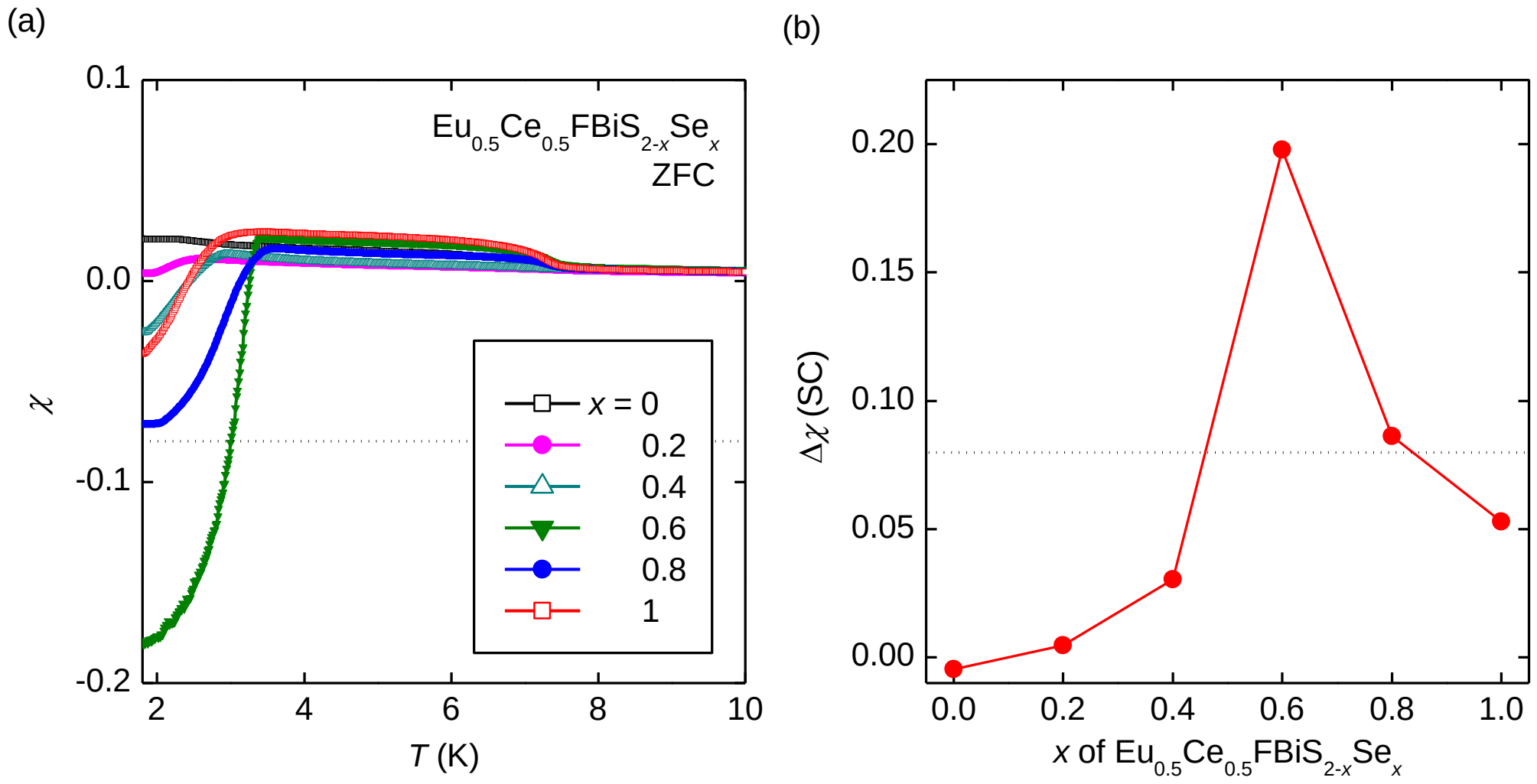

**Figure 5.** (a) Temperature ($T$) dependences of magnetic susceptibility ($\chi$) for $Eu_{0.5}Ce_{0.5}FBiS_{2-x}Se_x$ measured after zero-field cooling (ZFC). Dashed line represents the shielding fraction corresponding to -1/4$\pi$. $\chi$–$T$ plots after field cooling (FC) were shown in Figure S2. (b) Diamagnetic signal due to superconductivity ($\Delta\chi$(SC)) for $Eu_{0.5}Ce_{0.5}FBiS_{2-x}Se_x$. $\Delta\chi$(SC) was evaluated from the difference of $\chi$ between 4 K and 2 K.

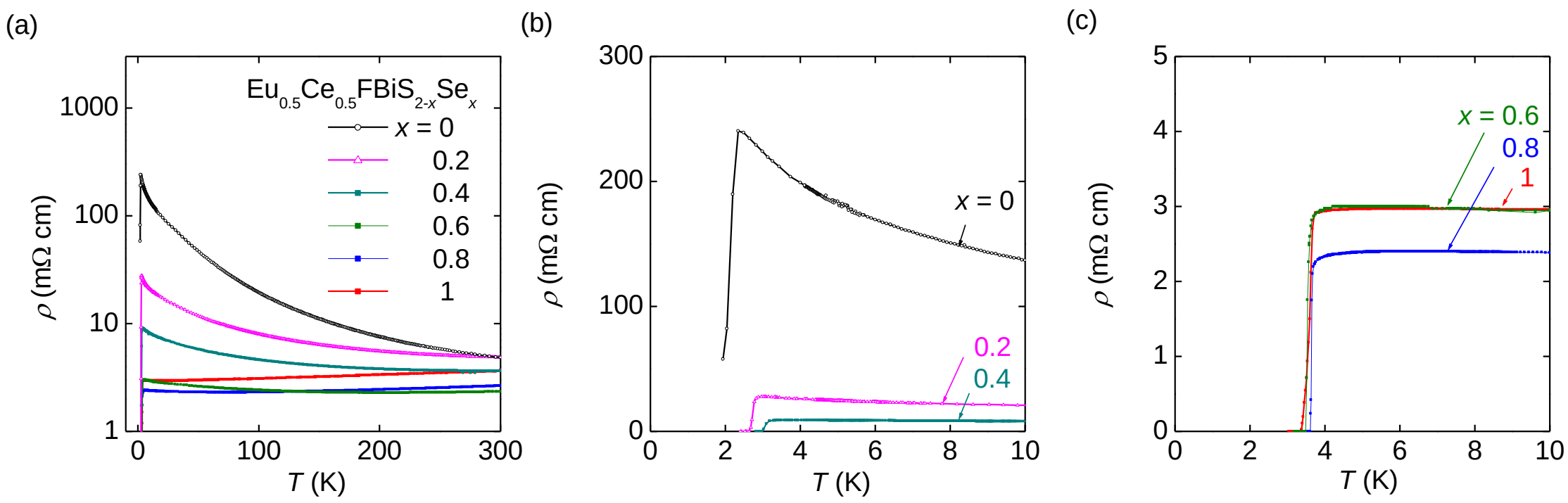

**Figure 6.** (a) Temperature ($T$) dependences of electrical resistivity ($\rho$) for $Eu_{0.5}Ce_{0.5}FBiS_{2-x}Se_x$. (b, c) $\rho$–$T$ plots below 10 K.

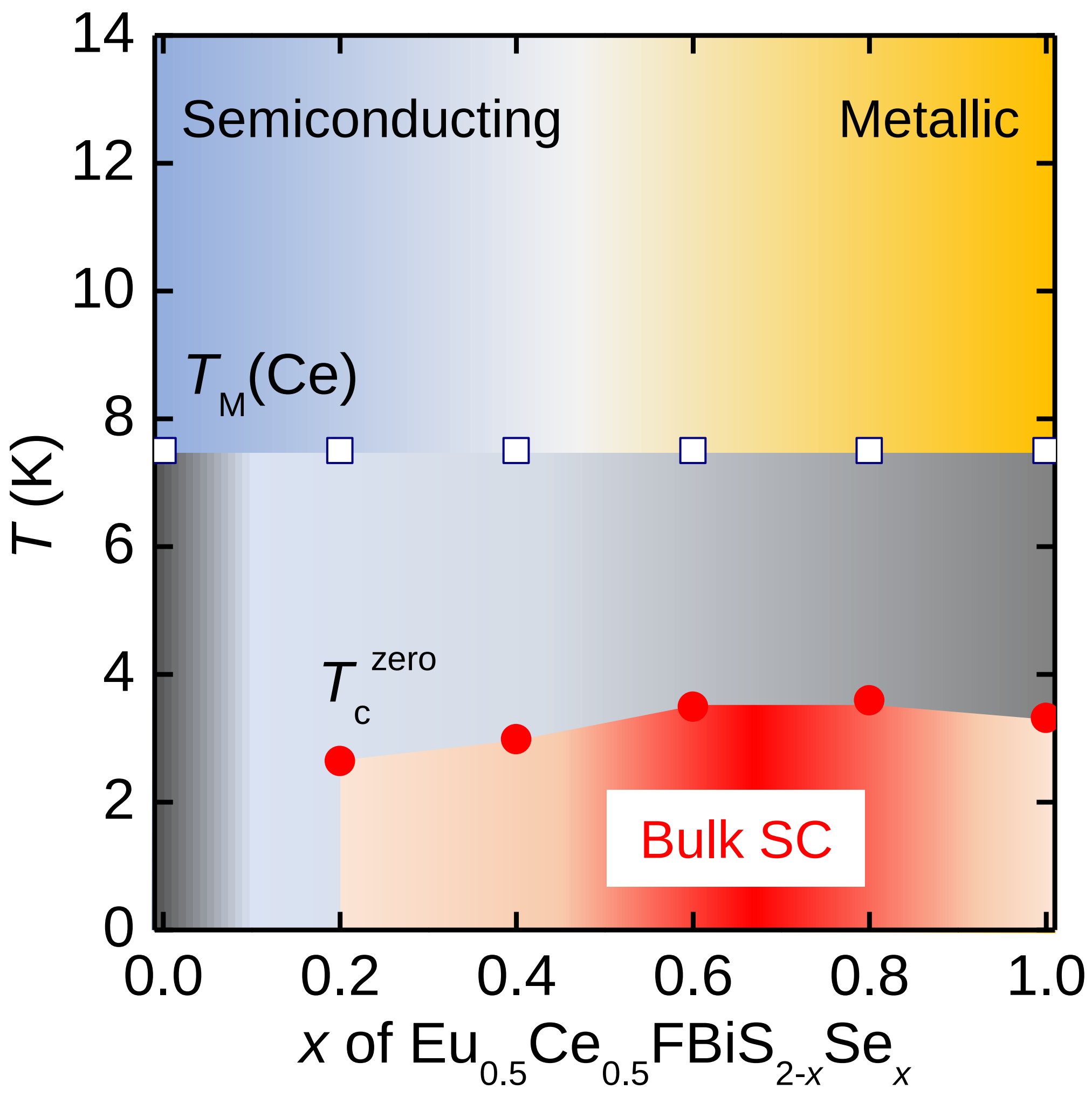

**Figure 7.** Superconductivity phase diagram of $Eu_{0.5}Ce_{0.5}FBiS_{2-x}Se_x$. $T_c^{zero}$ estimated from the $\rho$–$T$ data were plotted for $x$ = 0.2–1. SC denotes superconductivity.

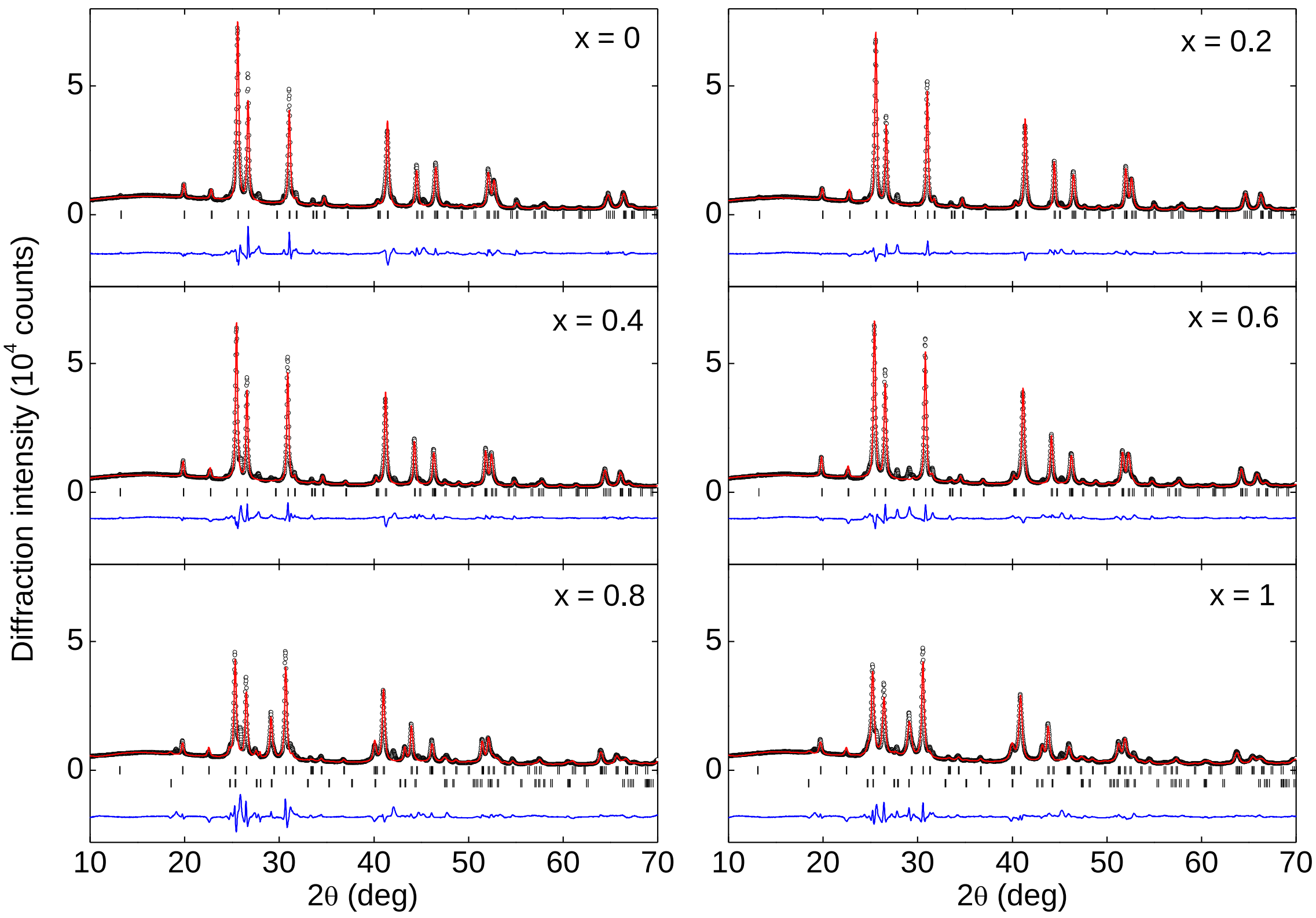

**Figure S1.** XRD patterns and Rietveld refinements of $Eu_{0.5}Ce_{0.5}FBiS_{2-x}Se_x$. For = 0.8 and 1, $Bi_2Se_3$ was included as a secondary phase. Bragg diffraction angles of $Bi_2Se_3$ were denoted in lower tick marks.

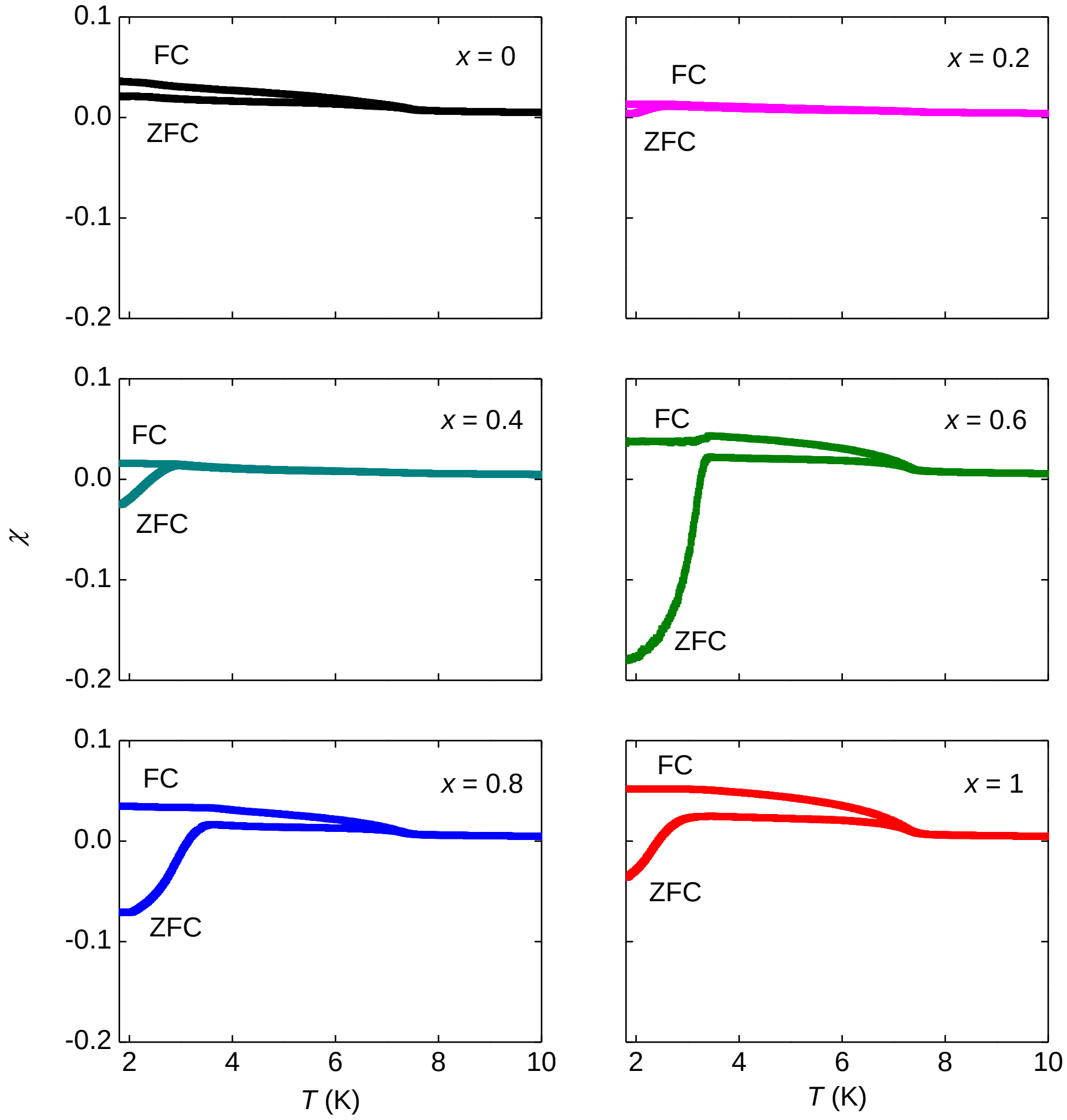

**Figure S2.** Temperature ($T$) dependences of magnetic susceptibility ($\chi$) for $Eu_{0.5}Ce_{0.5}FBiS_{2-x}Se_x$ measured after zero-field cooling (ZFC) and field cooling (FC). An gradual increase of $\chi$ below 8 K was most likely due to ferromagnetic order of $Ce^{3+}$ ions.